\documentclass[
reprint,
 amsmath,amssymb,
 aps,
prl,
]{revtex4-2}

\usepackage{graphicx}
\usepackage{dcolumn}
\usepackage{bm}
\usepackage{booktabs}
\usepackage[section]{placeins}
\usepackage{xcolor}
\usepackage{cancel}
\usepackage{ dsfont }
\usepackage{booktabs}
\usepackage{tabularray}
\usepackage{tablefootnote}
\usepackage{threeparttable}
\usepackage[normalem]{ulem}
\usepackage{bbm}
\usepackage{upgreek}
\newcommand{\bb}{{\mathbf{B}}}
\newcommand{\bB}{{\mathbf{b}}}

\begin{document}

\title{Topological invariant of periodic many body wavefunction from charge pumping simulation}

\author{Haoxiang Chen}
\affiliation{School of physics, Peking University, Beijing, China}
\affiliation{ByteDance Seed}

\author{Yubing Qian}
\affiliation{School of physics, Peking University, Beijing, China}
\affiliation{ByteDance Seed}

\author{Weiluo Ren}
\affiliation{ByteDance Seed}

\author{Xiang Li}
\affiliation{ByteDance Seed}
\email{lixiang.62770689@bytedance.com}

\author{Ji Chen}
\affiliation{School of physics, Peking University, Beijing, China}
\email{ji.chen@pku.edu.cn}


\begin{abstract}
Many-body topological quantum states host exotic quantum phenomena and lie at the forefront of developing next-generation quantum technologies. 
Recently emerged neural network wavefunction methods have established themselves as a powerful computational framework for accessing these states, enabling the variational machine learning calculation of the system’s ground state wavefunction. 
However, reliable computation of topological invariants remains an open challenge 
when the whole deterministic energy spectrum is not available.
In this work, we introduce a robust approach to determining topological invariant based on simulating the charge pumping process, by monitoring the response of polarization upon flux insertion.
By applying this method, we accurately extract the Chern numbers for Abelian fractional Chern insulators. 
Our approach also enables the first neural-network-wavefunction-based identification of anomalous composite Fermi liquid states. 
Our work resolves a key bottleneck in applying neural network wavefunctions to correlated topological matter, and the method proposed is also generally applicable to other many-body approaches, thereby opening up new avenues for future research in this field.
\end{abstract}

\newcommand{\btheta}{{\boldsymbol{\uptheta}}}

\maketitle

\begin{figure*}
    \centering
    \vspace{-0.7cm}
    \includegraphics[width=1.8\columnwidth]{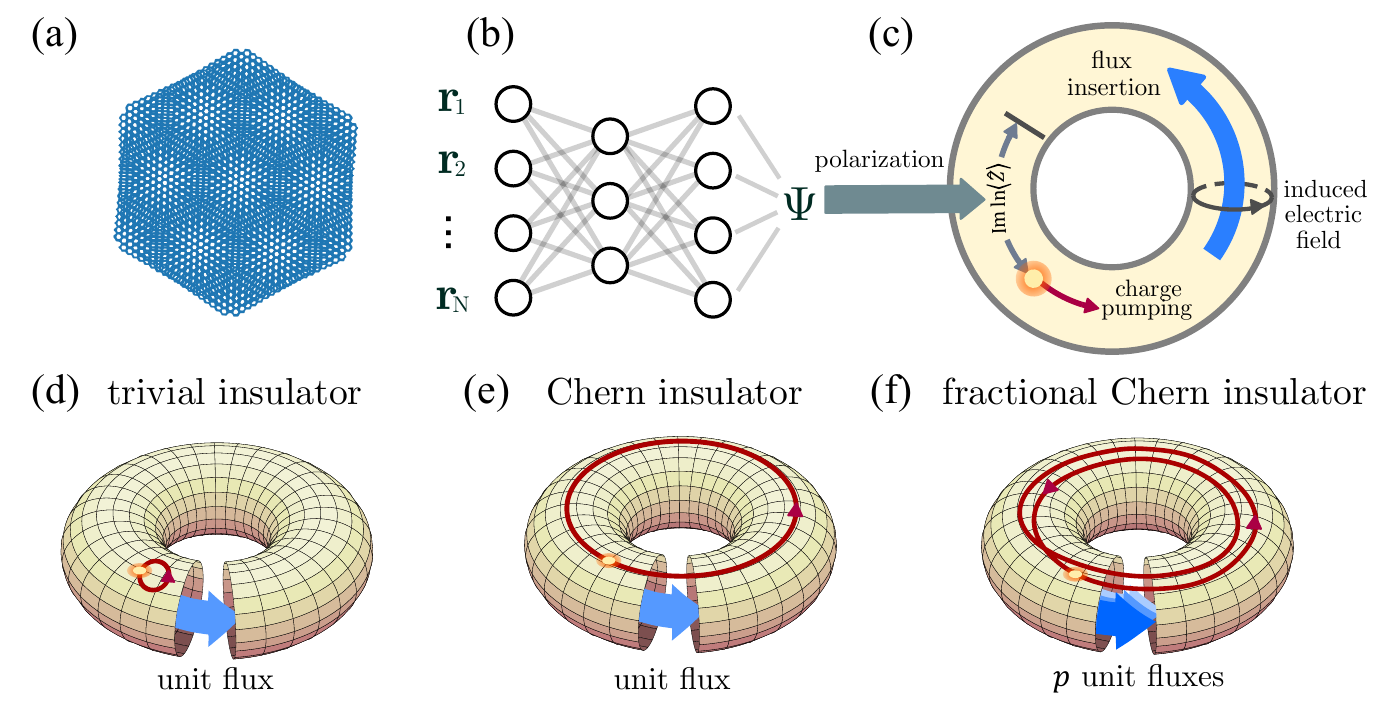}
    \vspace{-0.3cm}
    \caption{
        \label{Fig.ThoulessPumping}
        \textbf{Schematic of determining many-body Chern number with charge pumping.}
        (a) Schematic of moir\'e systems.
        (b) Schematic of continuous real space NNVMC, which maps electron coordinates to wavefunction value, 
        with first quantization scheme. 
        (c) Schematic of charge pumping, the yellow dot denotes the charge center defined with polarization operator.
        The blue arrow denotes the flux insertion inside the torus, resulting in twisted boundary condition along the meridian direction. The charge pumping, denoted by the red arrow, is perpendicular to the induced electric field.
        (d-f) Charge pumping for systems with different topological orders.
    }
\end{figure*}

The fractional quantum Hall effect is one of the most intriguing examples for correlated topological systems \cite{intro.FQHE,intro.FQHE2},
where the collective motion of electrons gives rise to fractionally quantized Hall conductance and anyonic statistics. 
Recently, the pursuit of correlated topological states has been revitalized in 
moir\'e systems, 
such as transition metal dichalcogenides \cite{intro.FCI1,intro.FCI2,intro.FCI3},
rhombohedral graphene-hBN \cite{intro.RhombohedralGraphene,intro.RhombohedralGraphene2,intro.RhombohedralGraphene3}
and twisted bilayer graphene \cite{intro.tBLG.FCI}
In moir\'e systems (Fig. \ref{Fig.ThoulessPumping}a), the synergistic interplay between strong many-body correlations and non-trivial quantum geometry stabilizes exotic quantum phases  \cite{intro.FQHE2FCI,intro.FQHE2FCI2,intro.FQHE2FCI3,intro.FQHE2FCI4,intro.FCI_ZeorChernBand}, 
most notably fractional Chern insulator (FCI) \cite{intro.theo.FCI,intro.theo.FCI2,intro.theo.FCI3,intro.theo.FCI4,intro.theo.FCI5,intro.theo.FCI6,FCI.theory} and composite Fermi liquid (CFL) \cite{intro.CFL,CFL.ED.Dong,CFL.ED.Reddy,FCI.ED} at zero magnetic field.
These novel states not only lay the cornerstone for future low-power topological electronic devices, 
but more attractively, the non-Abelian anyons they may host are ideal candidates for realizing topological quantum qubits.

From a numeric perspective, correlated topological states pose a formidable challenge for theoretical descriptions. 
The strong electronic correlations are indispensable and can not be reduced to effective single-particle pictures. 
Hence, the description must return to a many-body perspective, and the accurate computation of many-body wavefunctions is a longstanding challenge itself.
Recently, neural network variational Monte Carlo (NNVMC) emerges as a powerful method \cite{intro.NQS,intro.NQS2,intro.NQS3} to address this challenge, which exploits the expressive power of neural networks to represent the many-body wavefunctions (Fig. \ref{Fig.ThoulessPumping}b), 
and has achieved success in deriving the numerical many-body wavefunctions 
for quantum spin liquid~\cite{intro.QSL,intro.QSL2}, fractional quantum Hall states \cite{intro.FQHE+NN,intro.FQHE+NN2,intro.FQHE+NN3,intro.FQHE+NN4}, 
Hall crystals \cite{intro.QHC+NN}, as well as
FCIs and topologically trivial charge density waves (CDWs) \cite{FCI.NN,FCI.NN2,FCI.NN3-kspace,FCI.NN4}.
However, in regions where FCI states compete closely with topologically trivial and non-trivial charge density waves, accurately identifying the phase of neural network wavefunctions remains a challenge.
Therefore, the subsequent critical task lies in extracting reliable topological invariants from these numerical many-body wavefunctions.

Theoretically, the many-body Chern number $C$ is defined by the full space integral 
\begin{equation}
    \label{eq.c1}
    C = \int \frac{\text{d}^2{\btheta}}{2\pi\mathbbm{i}} 
    \left[
        \langle \partial_{\theta_y} {\Phi_\btheta} | \partial_{\theta_x} {\Phi_\btheta} \rangle
        - \langle \partial_{\theta_x} {\Phi_\btheta} | \partial_{\theta_y} {\Phi_\btheta} \rangle
    \right],
\end{equation}
where $|{\Phi_\btheta} \rangle$ is a periodic state obtained from $|{\Psi_\btheta}\rangle$ via a gauge transformation.
Here, $|{\Psi_\btheta}\rangle$ denotes the original many-body ground state wavefunction at twist boundary condition $\btheta=(\theta_x, \theta_y)$.
The transformation is given by $|{\Phi_\btheta} \rangle = \exp(-\frac{\mathbbm{i}}{2\pi} \sum_{a=x,y} \theta_a \bb_a \cdot\hat{\mathbf{r}} ) |{\Psi_\btheta}\rangle$, where $\hat{\mathbf{r}}$ is the electron position operator and $\bb_x,\bb_y$ are the supercell reciprocal lattice vectors.

To calculate the many-body Chern number via this definition, 
the effective Brillouin zone are divided into plaquettes, 
and the Berry curvature are computed for each plaquette using finite difference methods \cite{Chern.plaquette}. 
This calculation is usually done with exact diagonalization (ED) \cite{Non-Abelian.ED}, 
where wavefunctions and the overlap between them are available deterministically. 
Ref.~\cite{Chern.flat_berry_curvature} simplifies the calculation by introducing the one-plaquette formula, 
where the Berry curvature itself is proven to be virtually quantized and 
the whole integration in Eq.~\ref{eq.c1} can be replaced by calculating the integrand on a single plaquette. 
However, in the presence of ground state degeneracy, this method requires access to all degenerate ground states.
Alternatively, the Chern number can be determined from other global properties of wavefunction and without integration.
When the system respects certain rotational or inversion symmetry, 
the topological invariant can be obtained using wavefunctions only at high-symmetric points in the effective Brillouin zone \cite{Chern.symmetry,Z2.symmetry}. 
However, this approach is constrained by the system geometry and has not yet been extended to fractional Chern insulators.
Other methods draw upon the modern polarization theory \cite{Polarization,Polarization.Resta}, 
using Resta's polarization operator $\hat{Z}_{\bb}=\exp(\mathbbm{i}\bb\cdot\sum_i \hat{\mathbf{r}}_i)$ to determine the topological invariants. 
For instance, Ref.~\cite{Z2_marker} demonstrates that by using polarization along three specific directions, 
the topological invariant can be determined up to modulo 2. 
Another approach is adopted in Ref.~\cite{Pumping.iDMRG}, 
tracing the polarization phase shift when inserting magnetic flux.
The authors numerically calculated for the infinite cylinder geometry and also introduced the charge pumping for torus.
So far these polarization-based approaches are still confined to situations where topological invariants are integer.

Consequently, none of the methods mentioned above are readily compatible with neural network wavefunctions
when it comes to the FCI states.
To fully unleash the potential of neural networks on precision and scalability, 
especially its advantage of automatically taking band-mixing effect into account \cite{FCI.NN,FCI.NN2}, 
here we introduce a new method inspired by 
the concept of topological charge pumping \cite{TKNN,LaughlinArgument,Pumping.iDMRG},
to calculate the topological invariant for FCI states. 

We demonstrate that, despite the ground state degeneracy inherent to FCI states,  
the topological invariant can still be 
robustly extracted from the evolution of the polarization during charge pumping simulating.
As illustrated in Fig.~\ref{Fig.ThoulessPumping}~d-f,   
the phase shift of polarization at phase twist $\btheta=(\theta_x,\theta_y)$ is
    \begin{equation}
        \mathcal{C}(\theta_x) = 
        \frac{1}{2\pi} 
        \text{Im} \ln 
        \frac{\langle \Psi_{\theta_x,0} | \hat{Z}_{\bb_y} | \Psi_{\theta_x,0} \rangle }
            {\langle \Psi_{0, 0} | \hat{Z}_{\bb_y} | \Psi_{0, 0} \rangle }.
        \label{eq.ThoulessPumping}
    \end{equation}%
The wavefunction at twisted boundary condition $\btheta$ can be acquired by numerical simulation,
and the polarization phase is tracked for states with intermediate twists, and the continuous branch is chosen of the logarithm (see SI.~\textcolor{black}{\uppercase\expandafter{\romannumeral2}~A}).
When a unit flux is inserted, $\mathcal{C}(\theta_x=2\pi)$ is fractionally quantized, reflecting the topological invariant.
Here we evaluate the polarization along the supercell reciprocal lattice vectors $\bb_y$, 
chosen to ensure that $Z_{\bb_{y}}$ does not change the total momentum,
i.e., $N_e \bb_y = m\bB_x + n\bB_y$, 
where $N_e$ is the electron number, $m,n$ are integer numbers, 
and $\bB_x$, $\bB_y$ are primitive cell reciprocal basis vectors.
We also note that
the proposed method can be generalized to other twist directions (see SI.~\textcolor{black}{\uppercase\expandafter{\romannumeral3}~A})
and polarization operator with other wave vectors in the effective Brillouin zone (see SI.~\textcolor{black}{\uppercase\expandafter{\romannumeral3}~B}).

\begin{figure*}[htbp]
    \centering
    \includegraphics[width=2\columnwidth]{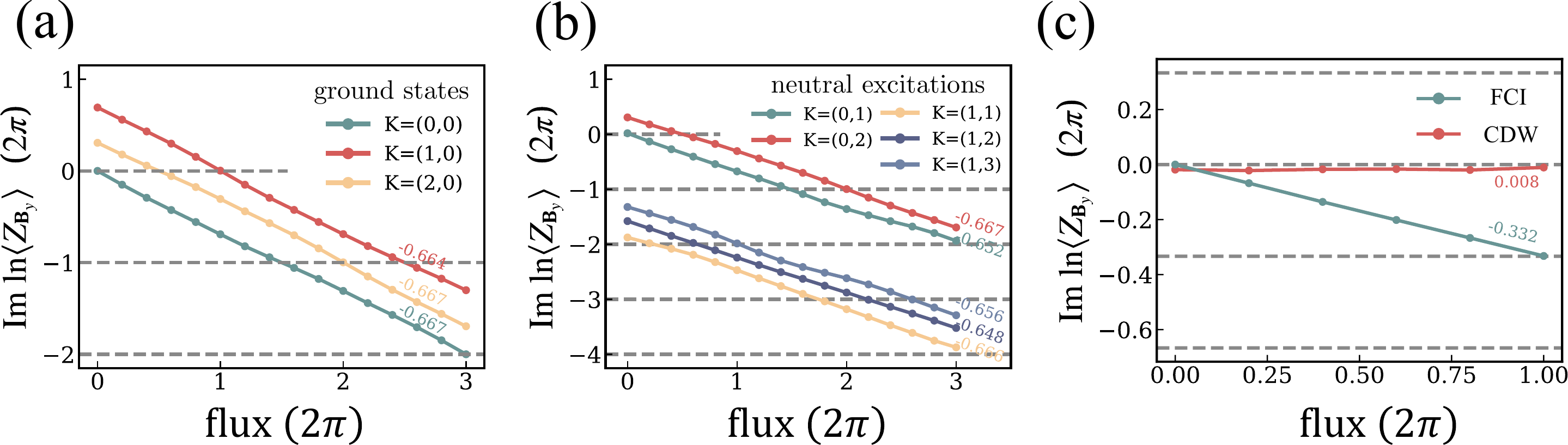}
    \caption{
        \label{Fig.FCI_and_CDW}
        \textbf{Chern number for fractional Chern insulators.}
        (a) Charge pumping for the nearly degenerate ground states at $2/3$ filling,  
        with total momentum as in label. 
        (b) Charge pumping for neutral excitations at $2/3$ filling.
        The $K=(1,1),~(1,2),~(1,3)$ states are shifted by $-2\pi$.
        (c) Charge pumping for $1/3$ filling for two geometries,
        with total momentum $K=(0,2)$ and $K=(0,0)$ for FCI and CDW state, respectively.
        The systems in (a) and (b) are calculated with $3\times4$ geometry.
        In (c), the CDW state is calculated with $3\sqrt{3}\times 3\sqrt{3}$ geometry,
        and the FCI state survives for $3\times4$ geometry that is imcompatible with the $\sqrt{3}\times\sqrt{3}$ CDW pattern.
        The numbers denote the slope of the lines.
    }
\end{figure*}

The phase shift $\mathcal{C}(\theta_x=2\pi)$ obtained in Eq.~\ref{eq.ThoulessPumping} can be justified as a reliable approximation of the Chern number $C$ via topological charge pumping. 
Following the generalized Laughlin's argument, 
the system's topological property can be determined with the pumped charge upon adiabatically inserting a unit flux.
As illustrated in Fig.~\ref{Fig.ThoulessPumping}c, consider the adiabatic
insertion of a magnetic flux along the $x$-direction in a toroidal geometry, 
the current is given by the time derivative of the polarization
$J=\text{d}P/\text{d}t=\frac{1}{2\pi}\text{d}~\text{Im} \ln \langle \hat{Z}_\bb \rangle / \text{d}t$,
according to the modern polarization theory \cite{Polarization,Polarization.Resta}.
The total pumped charge for unit flux insertion is thus $Q=\int \text{d}t J=\frac{1}{2\pi}\text{Im} \ln \langle\hat{Z}_\bb\rangle~|_{\theta=0}^{2\pi}$, 
Laughlin's argument states that tracing the flux insertion process enables the identification of whether the system is topologically non-trivial.
For trivial and integer quantum Hall states in Fig.~\ref{Fig.ThoulessPumping}d-e, 
the total pumped charge should be $0$ and $1$, and the phase shift of polarization is $0$ and $2\pi$, respectively.

For FCI states in Fig.~\ref{Fig.ThoulessPumping}f, the ground states are nearly-degenerate, 
and unit flux insertion does not guarantee the ground state return to the initial states, 
for which the charge center do not return to its initial position at zero flux and the pumped charge is non-integer. 
However the connection between polarization phase and topological invariant stays valid.
An intuition can be acquired from spectral flow, 
where the nearly-degenerate FCI ground states flow into each other upon flux insertion \cite{FCI.theory,FCI.theory.degeneracy}.
With $q$ unit fluxes inserted, the ground state can go back to the initial one and the polarization phase should consequently shift by multiples of $2\pi$.
Averaging over this process, the polarization phase shift per unit flux insertion is $2\pi\frac{p}{q}$, which directly encodes the fractional topological order, where $p$ and $q$ are integers.
A detailed discussion of the theory is provided in SI. \textcolor{black}{\uppercase\expandafter{\romannumeral1}}.



\begin{figure*}[htbp]
    \centering
    \includegraphics[width=1.9\columnwidth]{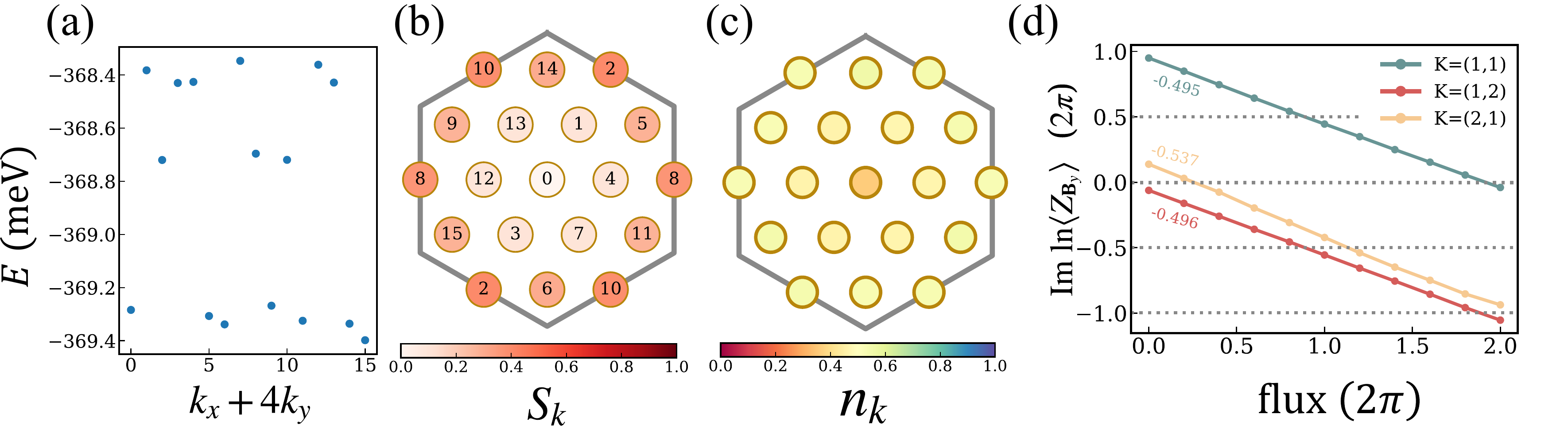}
    \caption{
        \label{Fig.EvenDenom}
        \textbf{Chern number for $1/2$ filling even denominator state.}
        (a) Momentum resolved spectrum. 
        (b) Structure factor, 
        the numbers are corresponding to the momentum sectors in (a).
        (c) Particle number momentum distribution.
        For (b) and (c), the results are averaged over the nearly degenerated ground states.
        (d) Charge pumping for ground state at 
        $K=(1,1),(1,2),(2,1)$ momentum sectors.
        All the calculations are performed at $1/2$ filling with $4\times 4$ geometry.
    }
\end{figure*}

To demonstrate the robustness of our method, we carry out calculations on the twisted $\text{MoTe}_2$ system, which has garnered substantial recent attention as a new platform for realizing the exotic quantum anomalous Hall states \cite{intro.FCI1,intro.FCI2,intro.FCI3,FCI.ED}.
Building on our prior NNVMC framework implemented in the DeepSolid package \cite{DeepSolid,FCI.NN}, we train neural network wavefunctions for a continuous model devised to capture the key electronic structure of this system.
The evaluation of polarization for neural network wavefunction is previously implemented in Refs.~\cite{DeepSolid,DeepSolid.Polarization} with stochastic integration.
More computational details are included in SI.~\textcolor{black}{\uppercase\expandafter{\romannumeral2}}.
We first validate our charge pumping calculation on
fully polarized states with
odd-denominator fractional hole fillings such as $2/3$ and $1/3$ (Fig. \ref{Fig.FCI_and_CDW}), where the neural network wavefunctions for these ground states are already established \cite{FCI.NN,FCI.NN2}.
We then extend our analysis to even-denominator fractional fillings, where we present the existence of 
CFL (Fig. \ref{Fig.EvenDenom}).

For the $2/3$ hole filling case, previous ED and NNVMC calculations have firmly identified the emergence of FCI ground state \cite{FCI.ED,FCI.NN,FCI.NN2}.
We calculated the charge pumping for all the three states in the low-energy subspace of a $3\times 4$ lattice, labeled by total momentum $K=(0,0), (1,0)$ and $(2,0)$.
The inserted flux is varied from $\theta=0$ to unit flux along the $y$-direction. 
The extracted numerical Chern number is $0.667, 0.664$ and $0.667$, respectively, in excellent agreement with the theoretical value of $2/3$.
Notably, the pumped charge is robust against variations in the flux insertion direction, as long as the flux is not parallel to the polarization direction (see SI.~\textcolor{black}{\uppercase\expandafter{\romannumeral3} B}).
Moreover, the results in Fig.~\ref{Fig.FCI_and_CDW}a is consistent with the spectral flow behavior in the FCI ground states. 
The polarization phases of these three states differ by $2\pi/3$, therefore upon inserting a unit flux, the $K=(0,0)$ flows into the $K=(2,0)$ state.
Consequently, the polarization center of $K=(0,0)$ state at a unit flux matches that of $K=(2,0)$ state at $0$ flux, with a corresponding polarization angle difference of $-4\pi/3$. 
In the SI, we show that the charge pumping methods can be extended to non-translational symmetry state (SI.~\textcolor{black}{\uppercase\expandafter{\romannumeral3} C}) and other Abelian FCI ground states, exemplified by the $3/5$ filling (SI.~\textcolor{black}{\uppercase\expandafter{\romannumeral3} D}).

Having established the efficacy of the charge pumping method for the ground state, we further probe its generality by testing charge pumping in neutral excited states across momentum sectors outside the ground-state subspace, with the corresponding results presented in Fig.~\ref{Fig.FCI_and_CDW}b.
While rigorous theoretical arguments and proofs for charge pumping have thus far been formulated exclusively for the ground state, we find that charge pumping behavior remains robust in neutral excited states.
This phenomenon is non-trivial and not intuitively obvious, but numerically our calculations demonstrate that fractional quantized pumping also emerges in neutral excited states, 
all the tested states show a similar slope near $-2/3$.
Other neutral excitation states ($K=(0,3),(2,3),(2,2),(2,1)$) are not demonstrated on Fig.~\ref{Fig.FCI_and_CDW}, but they can be related to $-K$ state by simultaneously inverting the layer and momentum, 
and the pumping slope should be the same for these states.

The $1/3$ filling is another case where Abelian FCI has been extensively discussed \cite{FCI.NN,FCI.NN2}.
Previous NNVMC calculations have pointed out an intricate competition between the FCI and the CDW state \cite{FCI.NN}.
Both geometry settings and band mixing effects are critical that can drive the ground state to entirely different states, and differentiating the two competing states also becomes highly challenging. 
And decisive evidences like particle entanglement spectrum are computationally demanding for NNVMC methods.
This underscores the necessity of an efficient and robust approach for evaluating topological invariants. 
As shown in Fig.~\ref{Fig.FCI_and_CDW}c, charge pumping approach serves as an effective means used to distinguish these two states at $1/3$ filling.
We focus our calculation on the $K=(0,2)$ and $K=(0,0)$ state for $3\times4$ and $3\sqrt{3}\times3\sqrt{3}$ systems respectively, 
which is compatible with FCI counting rule for both geometries. 
The CDW state with $3\sqrt3\times3\sqrt3$ geometry is previously confirmed in Refs.~\cite{FCI.NN,FCI.NN2}, 
and the charge pumping calculation yields an almost vanishing Chern number as visualized by a flat polarization curve.
In stark contrast, for the $3\times4$ geometry, 
which is incompatible with $\sqrt{3}\times\sqrt{3}$ CDW pattern,
the resulting numerical Chern number is $0.344$, in good agreement with the expected theoretical value of $1/3$.

Having established the robustness of the charge pumping approach for Abelian FCI states, 
we advance our investigation to the far less well-characterized CFL \cite{CFL.ED.Dong,CFL.ED.Reddy,CFL.ED.Reddy2} states at even-denominator fillings.
In the existing literature, the ground state at $1/2$ filling is conjectured to be a CFL state~\cite{CFL.ED.Dong}.
Calculating with a finite $4\times 4$ lattice, our NNVMC calculation yields ground states in momentum sectors that are approximately consistent with the CFL spectrum in ED calculation~\cite{CFL.ED.Dong} (Fig.~\ref{Fig.EvenDenom}a),
except for a previously unreported 
state at $K=(0,0)$ with a marginally lower energy.
Also, the ground states are featureless for the structure factor (Fig.~\ref{Fig.EvenDenom}b) and particle number distribution (Fig.~\ref{Fig.EvenDenom}c). 
In first Brillouin zone, the structure factor shows no peak  
and the electron numbers are all near $1/2$ without showing a Fermi surface, 
thus excluding the competing CDW and normal Fermi liquid phase.
Our charge pumping method can be applied to individual momentum sectors, 
and directly confirming the existence of non-trivial topological property.
As demonstrated in Fig.~\ref{Fig.EvenDenom}, 
a value of $\mathcal{C}=-0.495, -0.496, -0.537$ can be extracted via charge pumping in the $K=(1,1),(1,2),(2,1)$ sectors, which is consistent with the theoretical prediction.
Other CFL degenerate states at $K=(3,3),(3,2),(2,3)$ can be related by simultaneously inverting the layer and momentum 
and should behave similarly in pumping simulation.
It is worth noting that for a gapless system such as the CFL phase, the topological invariant is not well-defined. Nevertheless, the Hall conductance remains fractionally quantized, and our method remains applicable in cases where the polarization does not vanish, for example, in small systems.
These results exemplify the charge pumping as a robust identifier for the putative CFL from normal fermi liquid. 

To summarize, we demonstrate that the polarization operator
$\hat{Z}_\bb$,
when tracked under adiabatic flux insertion, serves as a robust estimator of topological invariants for fractional anomalous Hall states in strongly interacting flat-band systems.
We re-examine the Abelian FCI states in the twisted $\text{MoTe}_2$ system, and further extend the applicability of this method to CFL states. 
Notably, the charge pumping process can also be generalized to probe the whole anyonic statistics.
When choosing other wave vector $b$ in the Brillouin zone, 
the polarization operator $\hat{Z}_\bb$ generally maps one ground state to other within the manifold. 
The full anyonic statistics is encoded in the matrix representation of $\hat{Z}_\bb$, while in this work we only calculated for the diagonal component.
For future works, the mixed estimator of polarization, 
along with adiabatic flux insertion, 
can be a direct probe to anyonic statistics.


\begin{acknowledgments}
\noindent\textbf{Acknowledgments.} 
This work was supported by the National Science Foundation of China under Grant No. 12334003.
Y. Q. acknowledges supports from the National Natural Science Foundation of China under Grant No. 125B2083.
We also thank members of the ByteDance Seed Group for their inspiring ideas and encouragement.
\end{acknowledgments}

\bibliography{ref}

\end{document}



\title{Supplementary materials for \textit{Topological invariant of periodic many body wavefunction from charge pumping simulation}}

\author{Haoxiang Chen}
\affiliation{School of physics, Peking University, Beijing, China}
\affiliation{ByteDance Seed}

\author{Yubing Qian}
\affiliation{School of physics, Peking University, Beijing, China}
\affiliation{ByteDance Seed}

\author{Weiluo Ren}
\affiliation{ByteDance Seed}

\author{Xiang Li}
\affiliation{ByteDance Seed}
\email{lixiang.62770689@bytedance.com}

\author{Ji Chen}
\affiliation{School of physics, Peking University, Beijing, China}
\email{ji.chen@pku.edu.cn}


\renewcommand{\thefigure}{S\arabic{figure}}
\renewcommand{\theequation}{S\arabic{equation}}
\newcommand{\bx}{\mathbf{x}}
\newcommand{\bg}{\mathbf{g}}
\newcommand{\bG}{\bold{G}}
\newcommand{\bl}{\bold{l}}
\newcommand{\bk}{\bold{k}}
\newcommand{\br}{\bold{r}}
\newcommand{\bR}{\bold{R}}
\newcommand{\bp}{\bold{p}}
\newcommand{\bb}{\bold{B}}
\newcommand{\bB}{\bold{b}}
\newcommand{\bom}{\bold{\Omega}}
\newcommand{\balpha}{\boldsymbol{\upalpha}}
\newcommand{\btheta}{{\boldsymbol{\uptheta}}}
\newcommand{\Dx}{{\Delta\theta_x}}
\newcommand{\Dy}{{\Delta\theta_y}}
\newcommand{\bPsi}{{\boldsymbol{\Psi}}}
\newcommand{\bPhi}{{\boldsymbol{\Phi}}}

\maketitle

\section{Theories for charge pumping \label{SI.sec.prove_polarization}}

In the main text, a heuristic physical argument is given to justify 
the relationship between the polarization phase pumping and the topological invariant. 
In this section, we present a formal mathematical derivation that connects the charge pumping process to the many-body Chern number.
For systmes with integer topological invariants, 
a concise proof based on the path integral formulation has been given in Refs.~\cite{Pumping.iDMRG,SI.PolarizationIntegral}. We begin by offering a detailed derivation first, and then extend to fractional Chern insulators. 

Herein, we denote the ground state of $H=\hat{\bp}^2/2m+\hat{V}$ at phase-twist $\btheta=(\theta_x, \theta_y)$ as $|\Psi_\btheta\rangle$. 
Without loss of generality, we introduce phase-twist along the $x$-axis, 
and calculate the polarization $\hat{Z}_\bb$ along the $y$-axis by
choosing $\bb$ as $\bb_y$, the supercell reciprocal lattice vector along the $y$ direction.
The periodic part of $|\Psi_\btheta\rangle$ is denoted as $|\Phi_\btheta\rangle = \exp(-\mathbbm{i} \sum_{a=x,y}{{\theta_a} \hat{r}_a / L_a} ) |\Psi_\btheta \rangle$,
where $\hat{r}_a$ is the position operator and $L_a$ is the supercell lattice length. 
This periodic wavefunction $|\Phi_\btheta\rangle$ is the eigenstate of the gauge transformed Hamiltonian $\tilde{H}_\btheta = \sum_{a=x,y} \frac{1}{2m} (\hat{p}_a - \theta_a/L_a) + \hat{V}$.
The topological invariant can be derived from the Berry connection defined by $\Phi_\btheta$
\begin{equation}
    \begin{aligned}
        &  \mathcal{A}_\mu = \mathbbm{i} \langle \Phi_\btheta | \partial_{\theta_\mu} | \Phi_\btheta \rangle, 
        &  C = \frac{1}{2\pi} \oint  \text{d}\btheta \cdot \boldsymbol{\mathcal{A}}(\btheta).
    \end{aligned}
\end{equation}
We adopt a continuous gauge for $|\Phi_\btheta\rangle$ (and equivalently for $|\Psi_\btheta\rangle$) 
over the whole effective Brillouin zone.
For a non-trivial Chern band, this gauge must be multi-valued;
specifically, we impose boundary conditions
$|\Psi_{\theta_x=0,\theta_y}\rangle \neq |\Psi_{\theta_x=2\pi,\theta_y}\rangle$ 
and 
$|\Psi_{\theta_x,\theta_y=0}\rangle = |\Psi_{\theta_x,\theta_y=2\pi}\rangle$ 
.

With this gauge,
we first prove that the phase of polarization is related to the integral of the Berry connection
\begin{equation}
    \text{Im}~\text{ln} \langle \Psi_{\theta_x,0} |\hat{Z}_\bb| \Psi_{\theta_x,0} \rangle  = \int_{0}^{2\pi} \text{d}\theta_y \mathcal{A}_y(\btheta),
    \label{SI.eq.hook_Berry_connection_integral}
\end{equation}
Next, we demonstrate the accumulated polarization phase change is proportional to the Chern number, 
\begin{equation}
    \sum_{\theta_y=0}^{2\pi} 
        \big[
        \text{Im}~\text{ln} {\langle \Psi_{\theta_x+\Dx,0} | \hat{Z}_\bb | \Psi_{\theta_x+\Dx,0} \rangle}
        - \text{Im}~\text{ln} {\langle \Psi_{\theta_y,0} | \hat{Z}_\bb | \Psi_{\theta_x,0} \rangle} 
        \big]
        = 2\pi C
    \label{SI.eq.hook_Chern_calculation}
\end{equation}
Further, we extend Eq.~\ref{SI.eq.hook_Chern_calculation} to fractional Chern insulators.

To prove Eq.~\ref{SI.eq.hook_Berry_connection_integral}, 
we partition $\bb$ into $N$ pieces with $N\rightarrow \infty$.
Inserting complete basis $\sum_j |\Psi^j_{0,\Dy}\rangle \langle \Psi^j_{0,\Dy}|$ in between,
where $\Psi^j_{0,\theta_y}$ is the $j$-th eigenstate under twist angle $(0,\theta_y)$, 
the phase of polarization becomes
\begin{equation}
    \begin{aligned}
        &\text{Im}~\text{ln} \langle \Psi_{0,0}|\hat{Z}_{\bb}|\Psi_{0,0}\rangle \\
        =&\text{Im}~\text{ln} \sum_j \langle \Psi_{0,0} | \hat{Z}_{({\bb}/N)}^{N-1} | \Psi^j_{0,\Dy} \rangle 
        \langle \Psi^j_{0,\Dy} | \hat{Z}_{0,\Dy} | \Psi_{0,0} \rangle \\
        =&\text{Im}~\text{ln} \sum_j \langle \Psi_{0,0} | \hat{Z}_{({\bb}/N)}^{N-1} | \Psi^j_{0,\Dy} \rangle 
        \langle \Phi^j_{0,\Dy} | \Phi_{0,0} \rangle 
    \end{aligned},
    \label{SI.eq.def_phase}
\end{equation}
where $\Delta\theta_y=2\pi/N$, $|\Phi^j_\btheta\rangle = \exp(-\mathbbm{i} \sum_{a=x,y}{{\theta_a} \hat{r}_a / L_a} ) |\Psi^j_\btheta \rangle$.
$|\Phi^j_\btheta\rangle$ results from applying a gauge transformation from $|\Psi^j_\btheta\rangle$,
and is hence the $j$-th eigenstate for the gauge-transformed Hamiltonian
$\tilde{H}_\btheta = \sum_{a=x,y}{\frac{1}{2m}(\hat{p}_a-\theta_a/L_a)^2+\hat{V}}$.
For large enough $L_y$, $\tilde{H}_{0,\theta_y+\Dy}\approx \tilde{H}_{0,\theta_y} + \frac{\Delta\theta_x}{m L_x}{(p_y-\theta_y/L_y)}$,
and the ground state
$|\Phi_{0,\theta_y+\Dy}\rangle$ can be expanded using perturbation theory
\begin{equation}
    |\Phi_{0,\theta_y+\Dy}\rangle 
    \approx 
    e^{\mathbbm{i}\gamma} \left[
    |\Phi_{0,\theta_y}\rangle - \sum_{j\ne 0}\frac{\Dy}{m L_y}|\Phi^j_{0,\theta_y}\rangle\frac{\langle \Phi^j_{0,\theta_y}|\hat{p}_y-\theta_y/L_y| \Phi_{0,\theta_y}\rangle}{E^0 - E^j} 
    \right],
\end{equation}
where $\gamma$ is the geometric phase. 
Thus the overlap term in Eq.~\ref{SI.eq.def_phase} becomes
$\langle \Phi_{0,\theta_y+\Dy} |\Phi^j_{0,\theta_y} \rangle \propto \delta_j^0 + \mathcal{O}(\Dy L_y^{-1})$, 
and in Eq.~\ref{SI.eq.def_phase} we can consider only the ground state projection $|\Psi_{0,\theta_y}\rangle\langle \Psi_{0,\theta_y}|$ 
instead of summation over all complete basis set. 
By inserting $|\Psi_{\theta_x,0}\rangle\langle \Psi_{\theta_x,0}|$ in between every partition of $\bb$,
Eq.~\ref{SI.eq.def_phase} yields,
\begin{equation}
    \begin{aligned}
          \text{Im}~\text{ln} \langle \Psi_{0,0}|\hat{Z}_{\bb}|\Psi_{0,0}\rangle 
          &\approx { \text{Im}~\text{ln} \left[ \langle \Psi_{0,0}|\Psi_{0,2\pi}\rangle \prod_{\theta_y=\Dy}^{2\pi}\langle \Psi_{0,\theta_y} | {\hat{Z}_{\bb/N}}^N| \Psi_{0,\theta_y - \Dy} \rangle  \right]}\\
          &= { \text{Im}~\text{ln} \left[ \langle \Psi_{0,0}|\Psi_{0,2\pi}\rangle \prod_{\theta_x=\Delta \theta_x}^{2\pi}\langle \Phi_{0,\theta_y} | \Phi_{0,\theta_y - \Dy} \rangle  \right]}
    \end{aligned}
    \label{SI.eq.Wilson_loop}
\end{equation}
Expanding to the first order of $\Dy$,
and using $\ln(1+\epsilon) \approx \epsilon$, we obtain
\begin{equation}
    \begin{aligned}
        \text{Im}~\text{ln}{\langle \Phi_{0,\theta_y} | \Phi_{0,\theta_y-\Dy} \rangle } 
        &= \text{Im}~\text{ln}{\left[ 
            1 + \langle \Phi_{0,\theta_y} | \partial_{\theta_y} | \Phi_{0,\theta_y} \rangle (-\Dy)
        \right]} \\
        &\approx \mathbbm{i}{\langle \Phi_{0,\theta_y} | \partial_{\theta_y} | \Phi_{0,\theta_y} \rangle \Dy}
    \end{aligned},
    \label{SI.eq.ladder_to_Berry_connection}
\end{equation}
Combining Eq.~\ref{SI.eq.ladder_to_Berry_connection} and the continuous and multi-valued gauge $| \Psi_{\theta_x,0}\rangle = | \Psi_{\theta_x,2\pi} \rangle $,
Eq.~\ref{SI.eq.Wilson_loop} becomes
\begin{equation}
    \text{Im}\ln \langle \Psi_{0,0} | \hat{Z}_{\bb} | \Psi_{0,0} \rangle 
    \approx \mathbbm{i} \sum_{\theta_y=\Dy}^{2\pi} { \langle \Phi_{0,\theta_y} | \partial_{\theta_y} | \Phi_{0,\theta_y} \rangle} \Dy
    \approx \int_0^{2\pi} \text{d}\theta_y \mathcal{A}_y(\btheta).
\end{equation}
And for $\theta_x\neq 0$, this derivation stays valid,
thus proving Eq.~\ref{SI.eq.hook_Berry_connection_integral}.

Using Eq.~\ref{SI.eq.hook_Berry_connection_integral}, now we derive Eq.~\ref{SI.eq.hook_Chern_calculation}.
For this specific gauge, the $y$ component of Berry connection cancels out at $\theta_y=0$ and $\theta_y=2\pi$,
i.e., $\mathcal{A}_x(2\pi,\theta_y) - \mathcal{A}_x(0,\theta_y)=0$.
Accumulating the phase change over $\theta_y$ from $0$ to $2\pi$ gives
\begin{equation}
    \begin{aligned}
        \mathcal{C}(\theta_x=2\pi) \equiv & 
        \frac{1}{2\pi} \sum_{\theta_x=0}^{2\pi}{\text{Im}\ln \frac{\langle \Psi_{\theta_x +\Dx,0} | \hat{Z}_\bb | \Psi_{\theta_x+\Dx,0}\rangle }
                                                {\langle \Psi_{ \theta_x,0}  | \hat{Z}_\bb | \Psi_{ \theta_x,0} \rangle }}  \\
        \approx & \; \frac{1}{2\pi} \int \text{d}\theta_y~ \sum_{\theta_x=0}^{2\pi}[\mathcal{A}_y(\theta_x+\Dx,\theta_y) - \mathcal{A}_y(\theta_x,\theta_y)] \\
        = & \frac{1}{2\pi} \int \text{d}\theta_x~ [\mathcal{A}_x(0,\theta_y) - \mathcal{A}_x(2\pi,\theta_y)] 
            +\frac{1}{2\pi} \int \text{d}\theta_y~ [\mathcal{A}_y(2\pi,\theta_y) - \mathcal{A}_y(0,\theta_y)] \\
        = & \frac{1}{2\pi} \oint  \text{d}\btheta \cdot \boldsymbol{\mathcal{A}}(\btheta) = C \\
    \end{aligned}
    \label{SI.eq.Chern}
\end{equation}
This justifies Eq.~\ref{SI.eq.hook_Chern_calculation}.

Next, we demonstrate the pumping of polarization phase can also reveal the Chern number for fractional Chern insulators,
where the ground state is $D$-fold degenerate.
Consequently, the Berry connection and curvature are defined as matrices in the degenerate ground state subspace,
\begin{equation}
    \begin{aligned}
        \mathcal{A}_\mu^{mn} & = \mathbbm{i} \langle \Phi^{m} | \partial_{\theta_\mu} | \Phi^n \rangle \\
        \boldsymbol{\mathcal{F}} &= \partial_{\theta_x} \boldsymbol{\mathcal{A}}_y - \partial_{\theta_y} \boldsymbol{\mathcal{A}}_x 
        - \mathbbm{i} [\boldsymbol{\mathcal{A}}_x, \boldsymbol{\mathcal{A}}_y] 
    \end{aligned}
\end{equation}
where $m={0,1\cdots,D-1}$ denotes different ground states.
The Chern number is then
$
    C = \frac{1}{2\pi} \int \text{d}^2 \btheta ~ \text{tr}\;\boldsymbol{\mathcal{F}} 
$.
The trace of the commutator vanishes due to $\text{tr} (\boldsymbol{\mathcal{A}}_x \boldsymbol{\mathcal{A}}_y) = \text{tr} (\boldsymbol{\mathcal{A}}_y \boldsymbol{\mathcal{A}}_x)$,
simplifying the Chern number calculation to loop integrals over diagonal elements of the Berry connection matrix. 
\begin{equation}
    \begin{aligned}
        C   &= \frac{1}{2\pi} \int \text{d}^2 \btheta ~ \text{tr}\left[\partial_{\theta_x} \boldsymbol{\mathcal{A}}_y - \partial_{\theta_y} \boldsymbol{\mathcal{A}}_x \right] \\
            &= \sum_{m} \frac{1}{2\pi} \int \text{d}^2 \btheta \left[ \partial_{\theta_x} \boldsymbol{\mathcal{A}}^{mm}_y - \partial_{\theta_y} \boldsymbol{\mathcal{A}}^{mm}_x \right] \\
            &= \sum_{m} \frac{1}{2\pi} \oint \text{d}\btheta \cdot \boldsymbol{\mathcal{A}}^{mm} (\btheta)
    \end{aligned}
    \label{SI.eq.Chern_loop}
\end{equation}
By changing $\theta_y$ from $0$ to $2D\pi$,
the pumping of polarization phase for a single state $|\Psi^0\rangle$ is
\begin{equation}
    \begin{aligned}
    \mathcal{C}(\theta_x=2D\pi) &\equiv 
    \frac{1}{2\pi} \sum_{\theta_x=0}^{2D\pi}{
        \text{Im}\ln 
        \frac{\langle \Psi^{0}_{\theta_x+\Dx,0} | \hat{Z}_{\bb} | \Psi^{0}_{\theta_x+\Dx,0} \rangle}
             {\langle \Psi^{0}_{\theta_x,0} | \hat{Z}_{\bb} | \Psi^{0}_{\theta_x,0} \rangle} 
    }
    \end{aligned}
    \label{SI.eq.hook_fractional_phase_pumping}
\end{equation}
To relate this equation to Eq.~\ref{SI.eq.Chern_loop},
we make use of the emanant center-of-mass translational symmetry
$
    |\Psi_{\theta_x, \theta_y}^{m+1}\rangle 
    \approx 
    e^{\mathbbm{i}\beta(\btheta)}|\Psi_{\theta_x+2\pi, \theta_y}^{m}\rangle,
$
where $\beta$ depends on the gauge choice. 
As a result, the polarization phase shift for a single state over $\theta_x \in [0,2D\pi]$ in Eq.~\ref{SI.eq.hook_fractional_phase_pumping}
can be replaced by the summation over all ground states within $\theta_x \in [0,2\pi]$.

\begin{equation}
    \begin{aligned}
    \mathcal{C}(\theta_x=2D\pi)    &\approx \frac{1}{2\pi} \sum_{m=0}^{D-1}
    \sum_{\theta_x=0}^{2\pi}{
        \text{Im}\ln 
        \frac{\langle \Psi^{m}_{\theta_x+\Dx,0} | \hat{Z}_{\bb} | \Psi^{m}_{\theta_x+\Dx,0} \rangle}
             {\langle \Psi^{m}_{\theta_x,0} | \hat{Z}_{\bb} | \Psi^{m}_{\theta_x,0} \rangle} 
    } \\
    &= \sum_{m=0}^{D-1} \frac{1}{2\pi} \oint \text{d}\btheta \cdot \boldsymbol{\mathcal{A}}^{mm}(\btheta)  
    \end{aligned}
    \label{SI.eq.fci_loop}
\end{equation}
where the last step is derived in a manner similar to the non-degenerate case.
Combining Eq.~\ref{SI.eq.fci_loop} and Eq.~\ref{SI.eq.Chern_loop}, 
we conclude that the polarization phase pumping of a single ground state is sufficient 
to reveal the topological invariant for a fractional Chern insulator.

We remark here, the phase of $\langle \Psi^{0}_{\btheta}|\hat{Z}_\bb |\Psi^{0}_{\btheta} \rangle$ 
is only well-defined if the norm is non-vanishing, 
which requires $| \Psi^{m}_{\btheta} \rangle$ and $\hat{Z}_\bb | \Psi^{m}_{\btheta}\rangle$ are in the same momentum sector,
i.e., $N_e \bb \equiv (0,0)~\text{mod}~(\bB_x,\bB_y)$, with $N_e$ being the electron number and $\bB$ being the primitive cell reciprocal wavevectors.
When this restriction is violated, 
off-diagonal elements of the polarization matrix, $\langle\Psi^{m}_{\btheta}|\hat{Z}_\bb|\Psi^{n}_{\btheta}\rangle / [{\langle \Psi^{m}_{\btheta}|\Psi^{m}_{\btheta} \rangle}^{1/2} {\langle \Psi^{n}_{\btheta}|\Psi^{n}_{\btheta} \rangle}^{1/2}]$,
become important. 
The polarization matrices with different $\bb$ directions are non-commutative,
whose structure encodes the anyonic statistics. 
A more powerful diagostic for topological order beyond Chern number may be available by tracking the pumping of polarization matrices.

\clearpage
\newpage
\section{Calculation details}
\subsection{Charge pumping simulation}
{
In the main text, we identify fractional Chern insulator (FCI) and composite Fermion liquid (CFL) states by tracking the phase of polarization $\langle \exp(\mathbbm{i}\bb\cdot \sum_i {\br_i}) \rangle$ as a function of adiabatic flux insertion.
The calculations are performed with
torus geometry with twisted boundary conditions, and the flux insertion is implemented by changing the phase-twist angle $\theta\in[0,2\pi]$ with discrete steps of $0.4\pi$.
The polarization phase exhibits an approximate linear dependence on $\theta$, and its slope directly yields the single-state Chern number. 
Here, $\bb$ is a supercell reciprocal lattice vector,
non-parallel to the phase-twist 
and satisfying $N_e \bb = m \bB_x + n \bB_y$, 
where $N_e$ is the electron number and $\bB_x, \bB_y$ are the reciprocal lattice vectors for the primitive cell.

We use the neural network variational Monte Carlo (NNVMC) to capture the lowest ground state in each momentum sector.
We first optimize the neural network at $\theta = 0$ to acquire high-quality initial state.
For each subsequent twist angles, 
we initialize the network with the parameters of the preceding phase-twist angle, then continue optimization under the new boundary condition.
This strategy enforces adiabatic continuity and accelerating convergence. 
The hyperparameters for these two processes are listed in Table~\ref{SI.table.Hyperparameters}, 
denoted as ``initial training'' and ``pumping'' separately. 

}

{
For the optimized ground states under each $\theta$, the polarization $\langle \exp(\mathbbm{i}\bb\cdot \sum_i {\br_i}) \rangle$ is sampled with importance sampling using the probability density $|\Psi(\mathrm{\br})|^2$, 
where $r_i$ denotes the electron position for each sampled electron configuration.
During the sampling, the neural network parameters remain fixed.
}

\begin{table}[H]
  \centering
\caption{
    \textbf{Hyperparameters used for calculation.}
}
\begin{tabular}{c|c|c|c}
\hline
\hline
Hyperparameter & Value & Hyperparameter & Value\\
\hline
Dimension of one-electron layer & 256 & Dimension of two-electron layer & 32 \\
Number of layers  & 4 & Learning rate (initial training) & 3e-3\\
Optimizer & KFAC & Learning rate (pumping) & 3e-4 \\
Learning rate decay & 1 & Learning rate delay & 1e4 \\
Damping & 3e-4 & Constrained norm of gradient & 1e-3 \\
Momentum of optimizer & 0.0 & Batch size & 4096 \\
Training steps (initial training) & 4e4 & Clipping window of gradient & 20 \\
Training steps (pumping) & 2e4 & MCMC steps between iteration & 20 \\
MCMC burn in & 1e3 & Target MCMC acceptance & 55\% \\
MCMC move width & 2e-2 & Number of inference steps & 5e3 \\ 
Precision & Float32 & Number of determinants & 1 \\
\hline
\hline
\end{tabular}
\label{SI.table.Hyperparameters}
\end{table}

\subsection{Continuum model}
We use the continuum model for the twisted homo-bilayer transition metal dichalcogenides system \cite{model.Wu,FCI.ED}, 
the non-interacting part is
\begin{equation}
\begin{gathered}
    H_{0}(\br)=
    \begin{pmatrix}
    \frac{(-\mathbbm{i}\nabla-\mathbf{K}_+)^2}{2m} + \Delta_b(\br) & \Delta_T(\br) \\
    \Delta^*_T(\br) & \frac{(-\mathbbm{i}\nabla-\mathbf{K}_-)^2}{2m}+\Delta_t(\br)
    \end{pmatrix},\\
    \Delta_{b/t}(\br)=-2V\sum_{i=1,3,5}\cos(\bB_i\cdot\br\pm\delta), \\
    \Delta_{T}(\br)=\omega(1+e^{\mathbbm{i}\bB_2\cdot\br}+e^{\mathbbm{i}\bB_3\cdot\br}),\\
    \bB_i=\frac{4\pi}{\sqrt{3}a_M}\left[\cos(\frac{\pi(i-1)}{3}),\sin(\frac{\pi(i-1)}{3})\right],\\  
    \mathbf{K}_+=\frac{\bB_1+\bB_2}{3},\ \mathbf{K}_-=\frac{\bB_1+\bB_6}{3},\\
\end{gathered}
\label{SI.eq.Hamiltonian}
\end{equation}
where $a_M$ is the mori\'e superlattice constant,
and $V,\delta,\omega$ are parameters related to real materials.
And the total Hamiltonian reads, 
\begin{equation}
    H = \sum_i H_0(\br_i) + \sum_{i< j}{\frac{1}{\epsilon|\br_i-\br_j|}}
\end{equation}
For all the calculations in the main text and SI, we use the parameters relevant to $t\text{MoTe}_2$~\cite{FCI.ED}. The parameters are listed in Table.~\ref{SI.table.ModelParameters}.
\begin{table}[hb]
    \caption{
        \textbf{Parameters of $t{\rm MoTe_2}$.}
    }
    \begin{tabular}{c|c|c|c|c|c|c}
        \hline
        \hline
        $a_0\ ({\rm nm})$ & $\theta$ &$m (m_e)$ & $V$ (meV) & $\omega$ (meV) & $\delta$ & $\epsilon$ \\
            \hline
        0.352 & $2^\circ$ & 0.62 & 11.2 & -13.3 & $-91^\circ$ & 10\\
        \hline
        \hline
    \end{tabular}
  \centering
\label{SI.table.ModelParameters}
\end{table}

\clearpage
\newpage
\section{Extended results}

\subsection{Twist direction \label{SI.sec.twist_direction}}
\begin{figure}
    \centering
    \includegraphics[width=0.8\columnwidth]{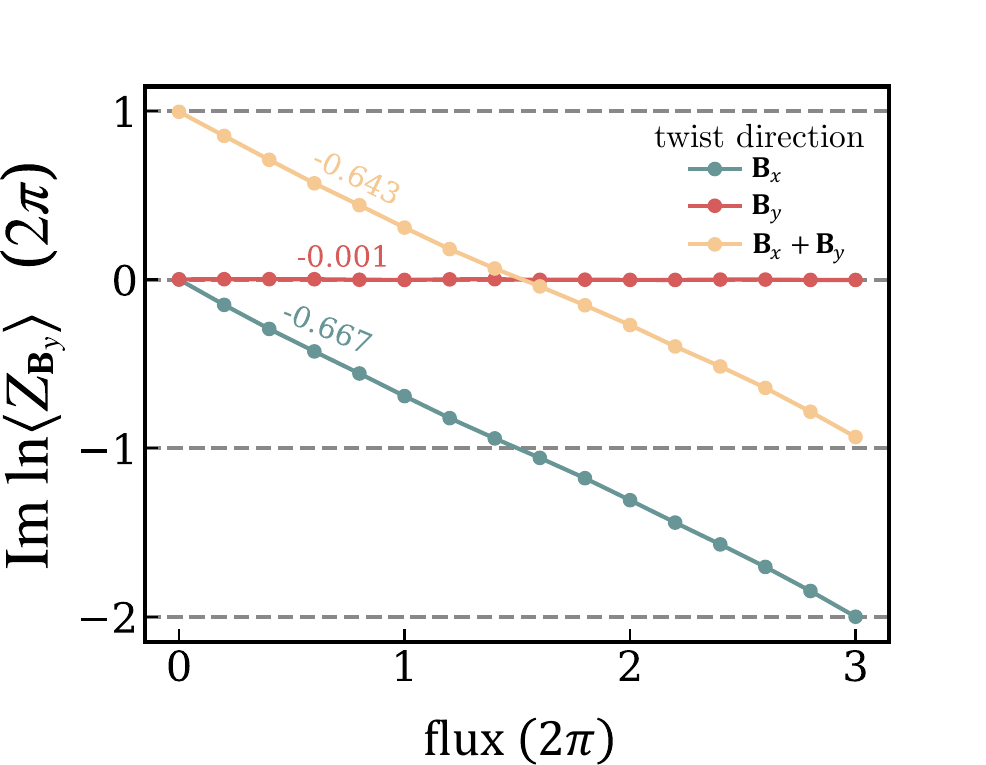}
    \caption{
        \label{SI.fig.twist_direction}
        \textbf{Effect of flux direction on charge pumping.}
        The polarization is calculated with $b_y$, 
        and $a_x, a_y$ are the supercell lattice vector respectively. 
        These states are calculated at $K=(0,0)$ momentum sector, with $3\times 4$ geometry. The numbers are the slope of each line.
    }
\end{figure}

We tested whether the charge pumping is stable for different flux directions.
The result is shown in Fig.~\ref{SI.fig.twist_direction}. 
For twist along $\bb_x$ and $\bb_x+\bb_y$, where the twist angle is not perpendicular 
to the polarization direction, 
the pumped charge is all close to $2/3$, compatible with the $2/3$ FCI state. 
On the contrary, no charge pumping occurs for twist along $\bb_y$, 
and the polarization phase is not changed. 

\clearpage
\newpage

\subsection{Charge pumping for other wave vectors \label{SI.sec.GenerlizedPumping}}
\begin{figure}
    \centering
    \includegraphics[width=0.8\columnwidth]{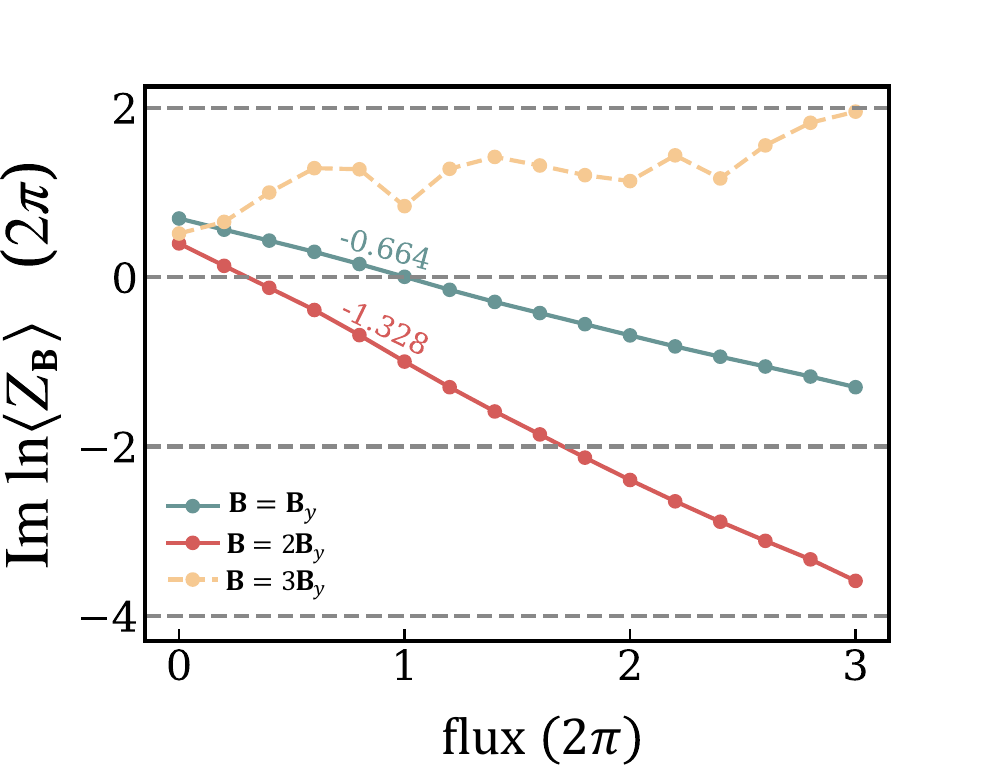}
    \caption{
        \label{SI.fig.other_wavevectors}
        \textbf{Charge pumping for various wave vectors.} 
        The calculation is performed with $3\times 4$ geometry in $K=(0,1)$ momentum sectors. 
        The numbers denote the slope of the corresponding line.
    }
\end{figure}

While in the main text, we perform charge pumping for only basis vectors of supercell reciprocal lattice, 
the charge pumping method with polarization operator can be generalized to other wave vectors as long as $N_e \bb = (0,0)~\text{mod}~(\bB_x, \bB_y)$. 
In Fig.~\ref{SI.fig.other_wavevectors}, when using wave vectors $\bb=2 \bb_y$ in the polarization operator, the charge pumping is also nearly linear, 
with a slope twice as large as the $\bb=\bb_y$ case. 
For larger wave vector $\bb=3 \bb_y$, the pumping is not quantized, 
because norm for the expectation $\langle Z_{3\bb_y} \rangle$ is very small, 
and the signal in polarization phase is lost in the Monte Carlo noise.

\clearpage
\newpage

\subsection{Topological invariant from non-translational-symmetric states.}
\begin{figure}
    \centering
    \includegraphics[width=0.8\columnwidth]{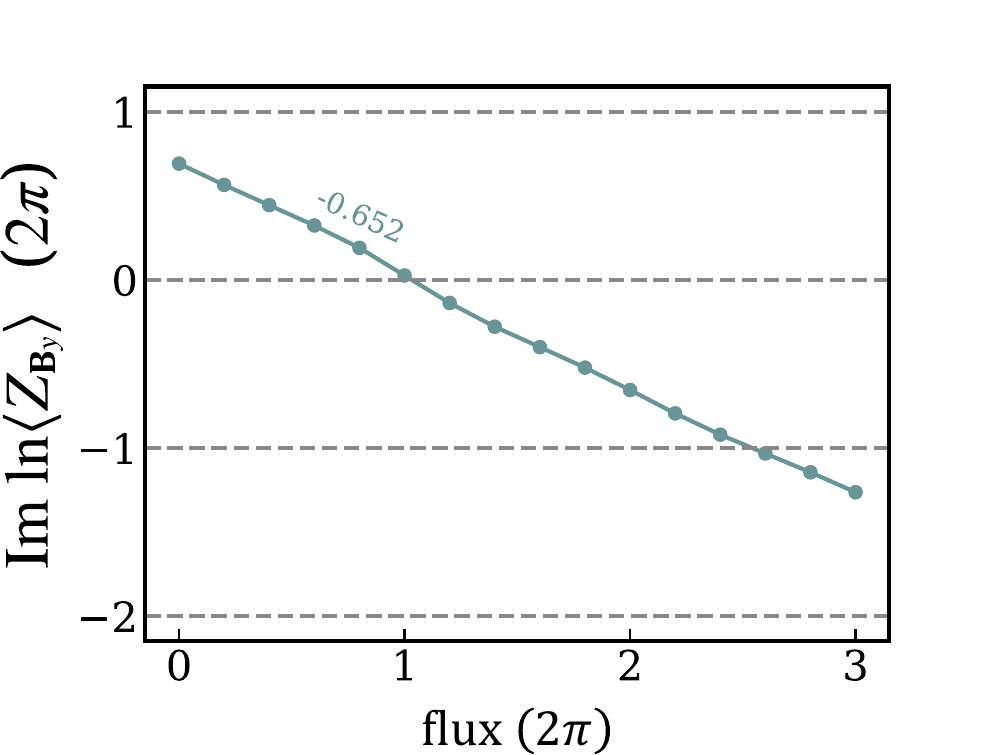}
    \caption{
        \label{Fig.NoKSymm}
        \textbf{Charge pumping for non-translational-symmetry state,}
        with $2/3$ filling at $3\times 4$ geometry.
        The polarization is chose along $b_y$ and the twist is along $x$ axis. The number denotes the slope of the line.
    }
\end{figure}

For NNQMC method, 
it's computationally more efficient to calculated approximate ground state
that is not eigenstates to the total momentum operator. 
These states, 
which are the superposition of momentum-resolved ground states, 
still lie in the low energy subspace spanned by topological states 
and carry essentially the same information for topological quantity.
We demonstrated in Fig.~\ref{Fig.NoKSymm}, 
the pumping charge is still roughly quantized to be $0.652$.

\clearpage
\newpage

\subsection{Charge pumping for $3/5$ filling \label{SI.sec.three-fifths}}
\begin{figure}
    \centering
    \includegraphics[width=0.8\columnwidth]{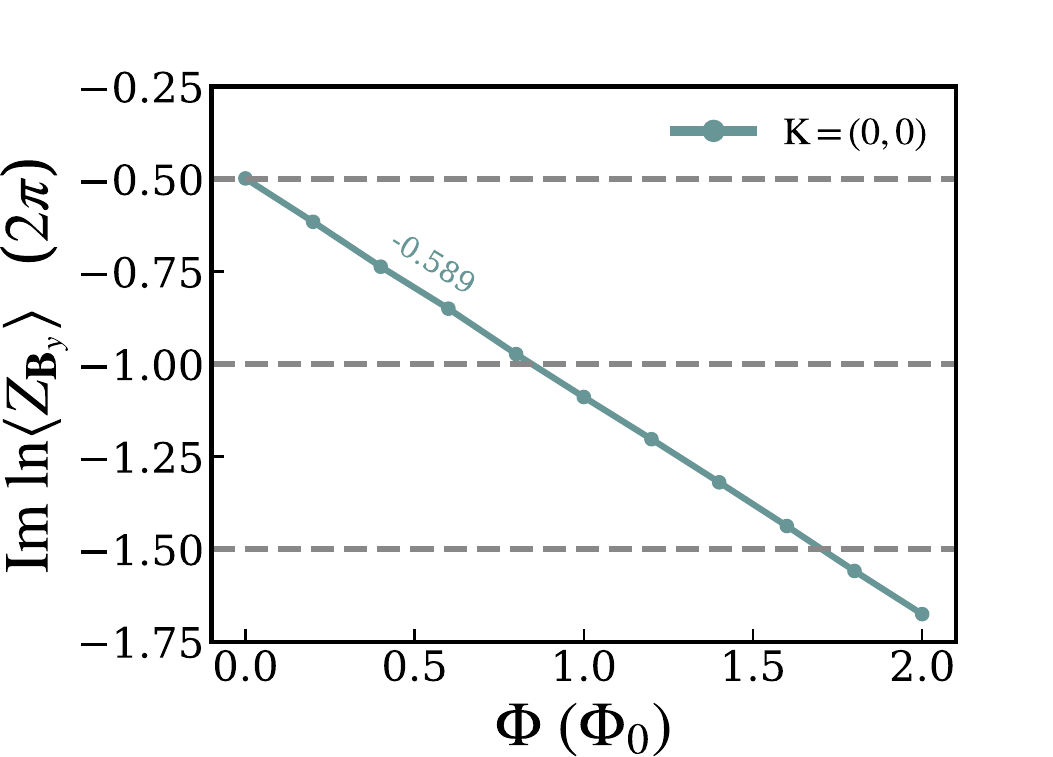}
    \caption{
        \label{SI.fig.Pumping.3/5}
        \textbf{Charge pumping for 3/5 state.}
        The calculation is performed with $K=(0,0)$ state and $5\times 3$ geometry.
    }
\end{figure}

Here we show the $3/5$ filling case as another example of the FCI state.
We calculated all ground states for a $5\times3$ geometry, and the energies are $-428.96,~-428.96,~-428.93$,
$~-428.92,~-428.95~\text{meV}$
for $\mathbf{K}=(0,0),~(1,0),~(2,0),~(3,0),~(4,0)$ states.
These states are nearly degenerate with energy difference much smaller than the gap to the excited state $\mathbf{K}=(0,1)$, whose energy is $-428.26~\text{meV}$.
This degeneracy is compatible with the counting rule for FCI states.
For the $\mathbf{K}=(0,0)$ state, 
the charge pumping is demonstrated in Fig.~\ref{SI.fig.Pumping.3/5}, 
the slope of the charge pumping is $-0.589$.

\bibliography{ref}